\documentclass[prb,preprint,eqsecnum,aps]{revtex4}
\usepackage{graphicx}

\newcommand{\e}{{\rm e}}

\newcommand{\x}{{\bf x}}

\newcommand{\br}{{\bf r}}
\newcommand{\s}{{\bf s}}

\newcommand{\bea}{\begin{eqnarray}}
\newcommand{\eea}{\end{eqnarray}}
\newcommand{\be}{\begin{equation}}
\newcommand{\ee}{\end{equation}}
\newcommand{\ba}{\begin{eqnarray}}
\newcommand{\ea}{\end{eqnarray}}

\newcommand{\nn}{\nonumber}
\newcommand{\la}{\label} 
\newcommand{\bv}{{\bf v}}

\newcommand{\m}{{\bf M}}

\begin{document} 
\title{A thousand fermions in a 3D harmonic trap via Monte Carlo simulations }
	
\author{Siu A. Chin }
	
\affiliation{Department of Physics and Astronomy, 
		Texas A\&M University, College Station, TX 77843, USA}
	
	%\date{\today}
\begin{abstract}
By use of a special wave function derived from similarly transformed propagators, this work shows that the energy of a thousand spin-balanced fermions in a three-dimensional harmonic potential can be accurately computed using the Monte Carlo method.

\end{abstract}

\maketitle

\section {Introduction}

It is well known that the sign problem has plagued the Monte Carlo simulations of many-fermion systems
in Path Integral\cite{cep96}, Diffusion\cite{ass07} and Ground State Path Integral\cite{sar00,yan17,chin20}
Monte Carlo methods. Even when there is no sign problem, as in the Variational Monte Carlo (VMC) method
with a trial wave function of the form
\be
\Psi={\det}_{\uparrow}|\phi_k(\br_i)|{\det}_{\downarrow}|\phi_k(\br_i)|\prod_{i<j}f(\br_{ij}),
\la{det2}
\ee 
where $k=\{n\ell m\}$ specifies a set of single particle states and $f(\br_{ij})$ is the Jastrow correlation function,
it remains technically burdensome to sample such a wave function for large number of fermions.  
For more than, say, 40 fermions, it is very tedious to specify the set of lowest energy single particle states, 
or to evaluate them in the Slater determinant ${\det}|\phi_k(\br_i)|$.
Except for atoms, most finite fermion systems use
the harmonic oscillator basis states. In this work, we show that there is a much simpler wave 
function for describing any number of non-interacting fermions in a harmonic
oscillator, with {\it no need of knowing the analytical form of its single particle wave functions}.  
As shown in Sect.\ref{newf}, the energy of up to a thousand spin-balanced fermions can be 
computed using this special wave function (\ref{detwf}), suggesting the feasibility of
doing very large scale VMC, or even Ground State Path Integral Monte Carlo\cite{sar00,yan17,chin20} 
calculations on 3D fermion systems.
  
The discovery of this wave function (\ref{detwf}) is rather circuitous. 
In order to account for the charge-density-wave (CDW) or Wigner crystal (WC)
density distributions observed in earlier calculations\cite{yos79,yos83}
on the fractional quantum Hall effect,  
Maki and Zotos\cite{mak83} postulated an ansatz wave function of the form
\be
\Psi(\x_1,\x_2\dots\x_N)\propto\det\Bigl|\exp[-\frac1{2}(\x_i-\s_j)^2]\Bigr|,
\la{det}
\ee
where each lowest Landau state's position $\x_i$ is localized at a variational position $\s_i$.
Unfortunately, this wave function's ground state energy at filling factor $\nu=1/3$
remains higher than that of Laughlin's wave function\cite{lau83}, even when
\{$\s_i$\} is the optimal two-dimensional triangular lattice.
However, this wave function remained useful for describing WC states in quantum dots
with and without a magnetic field\cite{kai02,har02}
when \{$\s_i$\} can be determined by minimizing the classical electron potential energy. 
 
For a given set of \{$\s_i$\}, the ansatz wave function
(\ref{det}) breaks translational and rotational invariance and remains {\it ad hoc} despite
the fact that symmetry-breaking density distributions similar to those produced by (\ref{det}) 
can be seen in unrestricted Hartree-Fock calculations\cite{yan99}. It was not until
recently that there is a natural way of deriving such symmetry-breaking wave functions
from similarly transformed propagators\cite{chin20}.

In this work we will first review how wave function such as (\ref{det}) can be derived in
Sect.\ref{stp}. In Sect.\ref{hfwf}, we verify analytically that the special wave function (\ref{detwf})
correctly gives wave functions for up to four free fermions in a $D$-dimension harmonic oscillator.
in Appendix A, we show that the special wave function (\ref{detwf})
gives the correct wave function for any number of fermions in a 1D harmonic oscillator.
In Sect.\ref{newf}, the Monte Carlo method is used to evaluate the ground state energy of up 
to a thousand spin-balanced fermions in a
3D harmonic oscillator. Conclusions are stated in Sect.\ref{con}.

\section {similarly transformed propagators}
\la{stp}

For completeness, we give here an expanded review below on similarly transformed propagators\cite{chin20}.
The key insight here is that the diagonal element of the imaginary time propagator
\be
G(\x,\x;\tau)=\langle \x|\e^{-\tau H}|\x\rangle,
\la{propa}
\ee
needed for extracting the ground state energy and wave function 
of a $D$-dimension Hamiltonian 
\be
H=-\frac12\nabla^2+V(\x)
\ee
at the large $\tau$ limit 
\be
\lim_{\tau\rightarrow\infty}G(\x,\x,\tau)\longrightarrow\psi_0^2(\x)\e^{-\tau E_0}+ \cdots,
\la{exg}
\ee
is invariant under the similarity transformation
\be
G(\x,\x;\tau)=\phi(\x)\langle \x|\e^{-\tau H}|\x\rangle\phi^{-1}(\x)
=\langle \x|\phi(\x)\e^{-\tau H}\phi^{-1}(\x)|\x\rangle
=\langle \x|\e^{-\tau\tilde H}|\x\rangle
\la{propb}
\ee
and can be computed equally from the transformed Hamiltonian\cite{chin90}
\be
\widetilde H\rho=\phi(\x)H\phi^{-1}(\x)\rho=-\frac12\nabla^2\rho+\nabla\cdot\Bigl[\bv(\x)\rho \Bigr]+E_L(\x)\rho,
\la{hp}
\ee
where the drift velocity $\bv(\x)$ and local energy $E_L(\x)$ are defined by
\be
\bv(\x)=\frac{\nabla\phi(\x)}{\phi(\x)}\qquad{\rm and}\qquad 
E_L(\x)=\frac{H\phi(\x)}{\phi(\x)},
\ee
provided that $\phi(\x)\ne 0$ at all $\x$.

As an example of this invariance, consider the case of the harmonic oscillator 
with $V(\x)=\x^2/2$.
The exact propagator for $H$ is known to be\cite{fey72}
\be
G(\x,\x_0;\tau)=\frac1{[2\pi\sinh(\tau)]^{D/2}}
\exp\left( -\frac1{2\sinh(\tau)}[ \cosh(\tau)(\x^2+\x_0^2)-2\x\cdot\x_0 ]\right),
\la{exaold}
\ee
while that of $\widetilde H$, with $\phi(\x)=\psi_0(\x)=\exp(-\x^2/2)$, is the
Ornstein-Uhlenbeck\cite{uhl} propagator
\be
\widetilde G(\x,\x_0;\tau)=\frac1{[2\pi T(\tau)]^{D/2}}
\exp\left[-\frac1{2T(\tau)}(\x-\x_0\e^{-\tau} )^2\right]\e^{-\tau E_0},
\la{fph}
\ee
where $T(\tau)=(1-\e^{-2\tau})/2$ and $E_0=D/2$. Despite their distinct appearances, they are
indeed the same when $\x_0=\x$:
\ba
G(\x,\x;\tau)&=&\frac1{[2\pi T(\tau)]^{D/2}}\e^{-\tau E_0}
\exp\left( -\frac{\x^2}{\sinh(\tau)}[ \cosh(\tau)-1 ]\right),\nn\\
&=&\frac1{[2\pi T(\tau)]^{D/2}}\e^{-\tau E_0}
\exp\left( -\frac{\x^2}{2T(\tau)}[ 1+\e^{-2\tau}-2\e^{-\tau} ]\right).\nn\\
&=&\widetilde G(\x,\x;\tau).
\la{geq}
\ea

The reason why one should consider the transformed propagator is that
while the exact propagator (\ref{propb}) can be computed using $H$ or $\widetilde H$
as above, their low order (in $\tau$) approximate propagators can be very different.
The first-order propagator computed from $H$ and $\widetilde H$ are respectively
\ba
G_1(\x,\x_0;\tau)&=&\frac1{[2\pi\tau]^{D/2}}
\exp\left[-\frac1{2\tau}(\x-\x_0)^2\right]\e^{-\tau V(\x_0)},
\la{pg1}\\
\widetilde G_1(\x,\x_0;\tau)
&=&\frac1{[2\pi\tau]^{D/2}}
\exp\left[-\frac1{2\tau}(\x-\x(\tau))^2\right]\e^{-\tau E_L(\x_0)},
\la{pg2}
\ea
where $\x(\tau)$ is the trajectory satisfying the drift equation
\be
\frac{d \x(\tau)}{d\tau}=\bv(\x(\tau))=\frac{\nabla\phi(\x(\tau))}{\phi(\x(\tau))}
\la{drift}
\ee
having the same initial position $\x(0)=\x_0$. 

In the case of the harmonic oscillator, (\ref{pg1}) bears no resemblance to (\ref{exaold}).
However, for $\widetilde H$ with $\phi(\x)=\psi_0(\x)=\exp(-\x^2/2)$, the
drift equation 
\be
\frac{d \x(\tau)}{d\tau}=-\x(\tau)
\la{dhar}
\ee
 yields the solution 
\be
\x(\tau)=\x_0\e^{-\tau},
\la{dsol}
\ee
resulting in the transformed first-order propagator 
\ba
\widetilde G_1(\x,\x_0;\tau)=\frac1{[2\pi\tau]^{D/2}}
\exp\left[-\frac1{2\tau}(\x-\x_0\e^{-\tau})^2\right]\e^{-\tau E_0}
\la{pg3}
\ea
which is already close to the exact propagator (\ref{fph}), differs only in
having the variance $\tau$ rather than $T(\tau)$.
As a matter of fact, its trace
\be
\int d\x\, \widetilde G_1(\x,\x;\tau)=[2\sinh(\tau/2)]^{-D}=\int d\x\, \widetilde G(\x,\x;\tau)=\int d\x\, G(\x,\x;\tau)
\ee
gives same partition function as the exact propagator (\ref{exaold}).
If one regards the {\it variance} $\tau$ in (\ref{pg3}) as a variational parameter 
fixed at $\tau=1$, then the (\ref{pg3}) will yield the correct wave function
(not its square) as $\tau\rightarrow\infty$ {\it in the trajectory equation} 
$\x(\tau)=\x_0\e^{-\tau}$.

This is of course an ideal case where one can take $\phi(\x)$ to be the
exact ground state $\psi_0(\x)$, but this illustrates the
possibility of improving the propagator by exploiting the invariance (\ref{propb}) under  
$\phi(\x)$. Historically, the transformed propagator (\ref{pg2}) with an approximate ground 
state $\phi(\x)$ has been used to accelerate the convergence of the Feynman-Kac path integral\cite{caf88}
and is the formal basis for doing Diffusion Monte Carlo with importance-sampling\cite{chin90,mos82,rey82}.

Generalizing to the case of $N$ {\it interacting} particles in a harmonic oscillator,
if one knows the exact bosonic ground state $\psi_0(\x_1,\x_2\cdots \x_N)$,
then an excellent approximation to the (unnormalized) single particle state would be
\ba
\psi_i(\x_i)=\exp\left[-\frac1{2}(\x_i-\s_i)^2\right],
\la{pg4}
\ea
where $\s_i=\x_i(\tau\rightarrow\infty)$. If $\s_i\ne 0$,
then it must be a {\it stationary} point of (\ref{drift}), implying that
\be
\nabla_{\x_i} \psi_0(\x_1,\x_2\cdots \x_N)|_{\s_i}=0.
\ee
This means that \{$\s_i$\} {\it is a set of particle positions that
 maximizes the bosonic ground state wave function, rather than just minimizes the classical
potential energy}. Such a discrete set of none zero \{$\s_i$\} would automatically breaks translational
and rotational symmetry even if $\psi_0(\x_1,\x_2\cdots \x_N)$ does not. This is
because one must necessarily start with a discreet set of initial positions 
\{$\x_{i0}$\}. The drift equation then evolves them toward
the set of discrete stationary points \{$\s_{i}$\}. Anti-symmetrizing the set of
single particle states (\ref{pg4}) then yields Maki-Zotos type wave function (\ref{det}).

A fundamental test of this transformed propagator approach of deriving
the many-fermion wave function is whether for 
$N$ {\it non-interacting  bosons} in a harmonic oscillator, the resulting wave function (\ref{det}) 
correctly describe $N$ {\it non-interacting  fermions} in the same potential.
We first verify analytically that this is indeed the case for up to four fermions in the next Section.

\section {Analytical few-fermion wave functions}
\la{hfwf}

For $N$-particles in a harmonic oscillator, the bosonic ground state wave function (ignoring normalization)
is
\be
\psi_B(\x_1,\x_2\cdots\x_N)=\prod_{i=1}^N\e^{-\x_i^2/2},
\ee
thereby yielding the same drift equation (\ref{dhar}) and the same solution (\ref{dsol}) for each 
particle: $\x_i(\tau)=\x_{i0}\e^{-\tau}$. Anti-symmetrizing the single particle state (\ref{pg4}) then
yields the $N$-fermion wave function as
\be
\Psi(\x_1,\x_2\dots\x_N)=\lim_{\s_j\rightarrow 0}\det\Bigl(\exp[-\frac1{2}(\x_i-\s_j)^2]\Bigr).
\la{detwf}
\ee
It is crucial to note here that the formalism requires anti-symmetrizing the single particle states 
first, before taking the $\s_j\rightarrow 0$ limit. Note also that
at a large finite $\tau$, $\s_j=\x_{j0}\e^{-\tau}$ is very close to zero, 
but never actually zero. Therefore analytically, the limit $\s_j\rightarrow 0$ means that one
should expand and keep only the leading non-vanishing powers of $\s_j$.

For $N=2$, the unnormalized antisymmetrized wave function is
\ba
\Psi(\x_1,\x_2)&=&
\det\left(\begin{array}{cc}
	\e^{-(\x_1-\s_1)^2/2} &  \e^{-(\x_1-\s_2)^2/2}\\
	\e^{-(\x_2-\s_1)^2/2} & \e^{-(\x_2-\s_2)^2/2}.
\end{array}\right)
\ea
Factoring out $\e^{-\x^2_1/2}$ and $\e^{-\x^2_2/2}$ from rows one and two
respectively and $\e^{-\s^2_1/2}$ and $\e^{-\s^2_2/2}$ from columns one and two
respectively gives
\ba
\Psi(\x_1,\x_2)
&=&	\lim_{\s_k\rightarrow 0} \e^{-(\x^2_1+\x^2_2+\s^2_1+\s^2_2)/2}
\det\left(\begin{array}{cc}
	\e^{\x_1\cdot\s_1} & \e^{\x_1\cdot\s_2}\\
	\e^{\x_2\cdot\s_1} & \e^{\x_2\cdot\s_2}
\end{array}\right),\la{twof}\\
&=&\e^{-(\x_1^2+\x_2^2)/2}\x_{21}\cdot\s_{21},
\la{tpwf}
\ea
where we have expanded to first order in $\s_j$ and defined $\x_{ij}=\x_i-\x_j$ and $\s_{ij}=\s_i-\s_j$.
This is the correct two-fermion wave function in a harmonic oscillator
of any dimension. For example, in three dimension, 
the two-fermion wave functions are three-fold degenerate given by
\ba
\Psi(\x_1,\x_2)\e^{(\x_1^2+\x_2^2)/2}&\propto&
\det\left(\begin{array}{cc}
	1 &  x_1\\
	1 &  x_2
\end{array}\right)
\quad{\rm or}\quad
\det\left(\begin{array}{cc}
	1 &  y_1\\
	1 &  y_2
\end{array}\right)
\quad{\rm or}\quad
\det\left(\begin{array}{cc}
	1 &  z_1\\
	1 &  z_2
\end{array}\right).
\ea
One sees that (\ref{tpwf}) is correctly a linear superposition of these three degenerate states
with arbitrary coefficients $(\s_2-\s_1)_k$. Note that one must have $\s_1\ne\s_2$, otherwise the wave function vanishes. 

Generalizing (\ref{twof}) to $N=3$ fermions gives
\ba
\Psi(\x_1,\x_2,\x_3)
&=&	\lim_{\s_k\rightarrow 0} \e^{-(\x^2_1+\x^2_2+\x^2_3)/2}
\det\left(\begin{array}{ccc}
	\e^{\x_1\cdot\s_1} & \e^{\x_1\cdot\s_2}& \e^{\x_1\cdot\s_3}\\
	\e^{\x_2\cdot\s_1} & \e^{\x_2\cdot\s_2}& \e^{\x_2\cdot\s_3}\\
	\e^{\x_3\cdot\s_1} & \e^{\x_3\cdot\s_2}& \e^{\x_3\cdot\s_3},
\end{array}\right). 
\ea
Multiply the first row by $\e^{(\x_2-\x_1)\cdot\s_1}=\e^{\x_{21}\cdot\s_1}$ and $\e^{\x_{31}\cdot\s_1}$ 
and subtract that from the second and third row respectively gives
\ba
&&\det\left(\begin{array}{ccc}
	\e^{\x_1\cdot\s_1} & \e^{\x_1\cdot\s_2}& \e^{\x_1\cdot\s_3}\\
	\e^{\x_2\cdot\s_1} & \e^{\x_2\cdot\s_2}& \e^{\x_2\cdot\s_3}\\
	\e^{\x_3\cdot\s_1} & \e^{\x_3\cdot\s_2}& \e^{\x_3\cdot\s_3}
\end{array}\right)
=
\det\left(\begin{array}{ccc}
	\e^{\x_1\cdot\s_1} & \e^{\x_1\cdot\s_2}& \e^{\x_1\cdot\s_3}\\
	0                  & \e^{\x_2\cdot\s_2}-\e^{\x_1\cdot\s_2+\x_{21}\cdot\s_1}&   
	\e^{\x_2\cdot\s_3}-\e^{\x_1\cdot\s_3+\x_{21}\cdot\s_1}\\
	0                  & \e^{\x_3\cdot\s_2}-\e^{\x_1\cdot\s_2+\x_{31}\cdot\s_1}& 
	\e^{\x_3\cdot\s_3}-\e^{\x_1\cdot\s_3+\x_{31}\cdot\s_1}
\end{array}\right)\nn\\
%&&\qquad\qquad\qquad\qquad=\e^{\x_1\cdot\s_1}
%\det\left(\begin{array}{cc}
%	\e^{\x_2\cdot\s_2}-\e^{\x_1\cdot\s_2+\x_{21}\cdot\s_1}& 
%	\e^{\x_2\cdot\s_3}-\e^{\x_1\cdot\s_3+\x_{21}\cdot\s_1}\\
%	\e^{\x_3\cdot\s_2}-\e^{\x_1\cdot\s_2+\x_{31}\cdot\s_1}& 
%	\e^{\x_3\cdot\s_3}-\e^{\x_1\cdot\s_3+\x_{31}\cdot\s_1}
%\end{array}\right)\nn\\
&&\qquad\qquad\qquad\qquad=\e^{\x_1\cdot\s_1}
\det\left(\begin{array}{cc}
	\e^{\x_2\cdot\s_2}(1-\e^{-\x_{21}\cdot\s_{21}})& 
	\e^{\x_2\cdot\s_3}(1-\e^{-\x_{21}\cdot\s_{31}})\\
	\e^{\x_3\cdot\s_2}(1-\e^{-\x_{31}\cdot\s_{21}})& 
	\e^{\x_3\cdot\s_3}(1-\e^{-\x_{31}\cdot\s_{31}}).
\end{array}\right)
\la{bed}
\ea
In the $\s_j\rightarrow 0$ limit, to second-order in $\s_j$, the above determinant becomes 
\ba
&=&
\det\left(\begin{array}{cc}
	(\x_{21}\cdot\s_{21})& 
	(\x_{21}\cdot\s_{31})\\
	(\x_{31}\cdot\s_{21})& 
	(\x_{31}\cdot\s_{31})
\end{array}\right)\nn\\
&=& (\x_{21}\cdot\s_{21})(\x_{31}\cdot\s_{31}) -(\x_{31}\cdot\s_{21})(\x_{21}\cdot\s_{31})\la{done}\\
&=&(\x_{21}\times\x_{31})\cdot(\s_{21}\times\s_{31})\la{exch}\\  
&=&(\x_{1}\times\x_{2}+\x_{2}\times\x_{3}+\x_{3}\times\x_{1})\cdot(\s_{1}\times\s_{2}+\s_{2}\times\s_{3}+\s_{3}\times\s_{1})
\la{sym}
\ea
The final form (\ref{sym}) shows that the cross-product $\x_{21}\times\x_{31}$ is anti-symmetric with any interchange $\x_i\leftrightarrow\x_j$ 
but symmetric in all $\x_i$.

In 3D, there are also three degenerate states for three fermions:
\ba
\det\left(\begin{array}{ccc}
	1 &   y_1 &z_1\\
	0 & y_{21}&z_{21}\\
	0 & y_{31}&z_{31}\\
\end{array}\right)
=y_{21}z_{31}-y_{31}z_{21}=(\x_{21}\times\x_{31})_x,
\ea
\ba
\det\left(\begin{array}{ccc}
	1 &   z_1 &x_1\\
	0 & z_{21}&x_{21}\\
	0 & z_{31}&x_{31}\\
\end{array}\right)
=z_{21}x_{31}-x_{31}z_{21}=(\x_{21}\times\x_{31})_y,
\ea
\ba
\det\left(\begin{array}{ccc}
	1 & x_1   &y_1\\
	0 & x_{21}&y_{21}\\
	0 & x_{31}&y_{31}\\
\end{array}\right)
=x_{21}y_{31}-x_{31}y_{21}=(\x_{21}\times\x_{31})_z.
\la{zstate}
\ea
The most general three-fermion wave function is therefore a linear combination of these 
three state as given by (\ref{exch}), with arbitrary coefficients $(\s_{21}\times\s_{31})_k$.

In 2D, only $(\s_{21}\times\s_{31})_z$ is possible and therefore the three-fermion wave function is give by (\ref{zstate}) only, with no degeneracy, because $N=3$ is a closed-shell state in 2D.

In 1D, (\ref{done}) vanishes, and one must expand (\ref{bed}) to the third order in $\s_j$
to obtain the 3-fermion wave function. In Appendix A,  
we derive this wave function and show that (\ref{detwf}) correctly produces the
general $N$-fermions wave function in 1D.

For $N=4$ fermions, similar steps in the limit of $\s_k\rightarrow 0$ gives
\ba
&&\det\left(\begin{array}{cccc}
	\e^{\x_1\cdot\s_1} & \e^{\x_1\cdot\s_2}& \e^{\x_1\cdot\s_3}& \e^{\x_1\cdot\s_4}\\
	\e^{\x_2\cdot\s_1} & \e^{\x_2\cdot\s_2}& \e^{\x_2\cdot\s_3}& \e^{\x_2\cdot\s_4}\\
	\e^{\x_3\cdot\s_1} & \e^{\x_3\cdot\s_2}& \e^{\x_3\cdot\s_3}& \e^{\x_3\cdot\s_4}\\
	\e^{\x_4\cdot\s_1} & \e^{\x_4\cdot\s_2}& \e^{\x_4\cdot\s_3}& \e^{\x_4\cdot\s_4}
\end{array}\right)
\rightarrow\det\left(\begin{array}{ccc}
	(\x_{21}\cdot\s_{21})&
	(\x_{21}\cdot\s_{31})& 
	(\x_{21}\cdot\s_{41})\\
	(\x_{31}\cdot\s_{21})& 
	(\x_{31}\cdot\s_{31})& 
	(\x_{31}\cdot\s_{41})\\
	(\x_{41}\cdot\s_{21})& 
	(\x_{41}\cdot\s_{31})& 
	(\x_{41}\cdot\s_{41})
\end{array}\right).
\la{fourc}
\ea
To evaluate this determinant, choose a coordinate system such that $\s_{21}$ is along 
the x-axis with $\s_{21}=(s_{21},0,0)$ and $\s_{31}$ at an angle $\theta\ne 0$ relative to $\s_{21}$
with $\s_{31}=(s_{31}\cos\theta,s_{31}\sin\theta,0)$ and $\s_{41}=(s_{41x},s_{41y},s_{41z})$. 
Multiply the $\s_{21}$ column in (\ref{fourc}) by $s_{31}\cos\theta/s_{21}$ 
and subtract it from the $\s_{31}$ column so that that column now has $\s_{31}=(0,s_{31}\sin\theta,0)$. 
Multiply the $\s_{21}$ column above by $s_{41x}/s_{21}$ and subtract it from $s_{41}$ column
column gives $\s_{41}=(0,s_{4y},s_{41z})$. Finally, multiply the $\s_{31}$ column by $s_{41y}/(s_{31}\sin\theta)$
and subtract it from $s_{41}$ column gives effectively $\s_4=(0,0,s_{41z})$. This then results 
\ba
\det({\cdots})
&=&[s_{21} s_{31}\sin\theta s_{41z}]\x_{21}\cdot(\x_{31}\times\x_{41})\nn\\
&=&[(\s_{21}\times \s_{31})\cdot \s_{41}]\x_{21}\cdot(\x_{31}\times\x_{41})\nn\\
&=&[\s_{21}\cdot(\s_{31}\times\s_{41})] \x_{21}\cdot(\x_{31}\times\x_{41}),
\la{fours}
\ea
which is exactly proportional to the closed-shell four-fermion wave function in 3D:
\be
\det\left(\begin{array}{cccc}
	1 & x_1&y_1&z_1\\
	1 & x_2&y_2&z_2\\
	1 & x_3&y_3&z_3\\
	1 & x_4&y_4&z_4\\
\end{array}\right)
=
\det\left(\begin{array}{cccc}
	1 & x_1&y_1&z_1\\
	0 & x_{21}&y_{21}&z_{21}\\
	0 & x_{31}&y_{31}&z_{31}\\
	0 & x_{41}&y_{41}&z_{41}\\
\end{array}\right)
=
\det\left(\begin{array}{ccc}
	x_{21}&y_{21}&z_{21}\\
	x_{31}&y_{31}&z_{31}\\
	x_{41}&y_{41}&z_{41}\\
\end{array}\right)
=\x_{21}\cdot(\x_{31}\times\x_{41}).
\la{wcor}
\ee
Since this closed-shell state is unique, it is multiplied only by a single coefficient 
$\s_{21}\cdot(\s_{31}\times\s_{41})$ corresponding to the volume formed by
any non-coplanar set of relative vectors $\s_{k1}$. 

Since the triple product $\s_{21}\cdot(\s_{31}\times\s_{41})$ vanishes in 2D,
one must expand to the fourth-order in $\s_i$ to obtain the
four-fermions wave function in 2D. We will not bother with this task here. 
However, for any dimension and any number of fermions,
one can always evaluate the wave function (\ref{detwf}) numerically.
In the next Section, we show that the wave function (\ref{detwf}) can be used in Monte Carlo calculations
to obtain the ground state energy of up to 1000 spin-balanced fermions in a 3D harmonic oscillators.

\section {Numerical evaluation of fermion energies}
\la{newf}

In this work, we will only focus on the fermion part of the wave function (\ref{det2}) and
leave the Jastrow correlation for later more specific applications.
The wave function can be rewritten as
\be
\Psi(\x_1,\x_2\dots\x_{2N})=\det\m_{\uparrow}(\x_1,\dots\x_N)\det\m_{\downarrow}(\x_{N+1},\dots\x_{N+L})=\exp(S_{\uparrow}+S_{\downarrow}),
\la{wfd}
\ee
with
\be
S_{\uparrow}(\x_1,\x_2\dots\x_{N})=\ln (\det \m_{\uparrow})
\ee
where $\m_{\uparrow}$ is the $N\times N$ matrix for particles $i=1$ to $N$
\be
M_{\uparrow ij}=\exp\left[-\frac1{2}(\x_i-\s_j)^2\right].
\la{mat}
\ee
Similarly one can define
$S_{\downarrow}=\ln (\det \m_{\downarrow})$ where $\m_{\downarrow}$ is the $L\times L$ matrix (\ref{mat}) 
for particles $i=N+1$ to $N+L$. The computer code evaluates the determinant of both $N\times N$ and $L\times L$
matrices and can study various unequal spin cases for $N\ne L$. 
Here, only results for the spin-balance case of $L=N$ is presented. The resulting energy is directly
computed from two determinants and not from doubling the energy of a single determinant. 

The local energy function for the spin-up fermions is then given by (See Appendix B)
\ba
E_L^{\uparrow}&=&\frac{H\Psi}{\Psi}=\frac{\sum_i^N[-\frac12\nabla_i+V(\x_i)]\Psi}{\Psi}\nn\\
&=&\sum_{i=1}^N\left[-\frac12[\nabla^2_i S_{\uparrow}+(\nabla_i S_{\uparrow})^2]+\frac12 \x_i^2\right],\nn\\
&=&N\frac{D}2-\frac12\sum_{i=1}^N(\s_i^2-\tilde\s_i^2 )
-\frac12\sum_{i=1}^N(\x_i-\tilde\s_i)^2 + \frac12\sum_{i=1}^N\x_i^2,
\ea
where
\be
\tilde\s_i=\sum_{k=1}^N \s_k M_{\uparrow ik} M^{-1}_{\uparrow ki}.
\ee
The spin-down local energy $E_L^{\downarrow}$ is similarly defined for particles $i=N+1$ to $2N$.
One then computes the energy expectation value
\be
E_{2N}=\frac{\int d\x_1\cdots d\x_{2N} (E_L^{\uparrow}+E_L^{\downarrow})\Psi^2(\x_1,\x_2\dots\x_{2N})}
             {\int d\x_1\cdots d\x_{2N}\Psi^2(\x_1,\x_2\dots\x_{2N})}
\ee
using the Metropolis {\it et al.} algorithm\cite{met}. Since one is sampling $\Psi^2$, 
there is no sign problem in this calculation. 
The fermion character of the problem is encapsulated in $\tilde\s_i$, which requires
computing the inverse matrix $\m^{-1}$.

\begin{figure}[hbt]
	\includegraphics[width=0.48\linewidth]{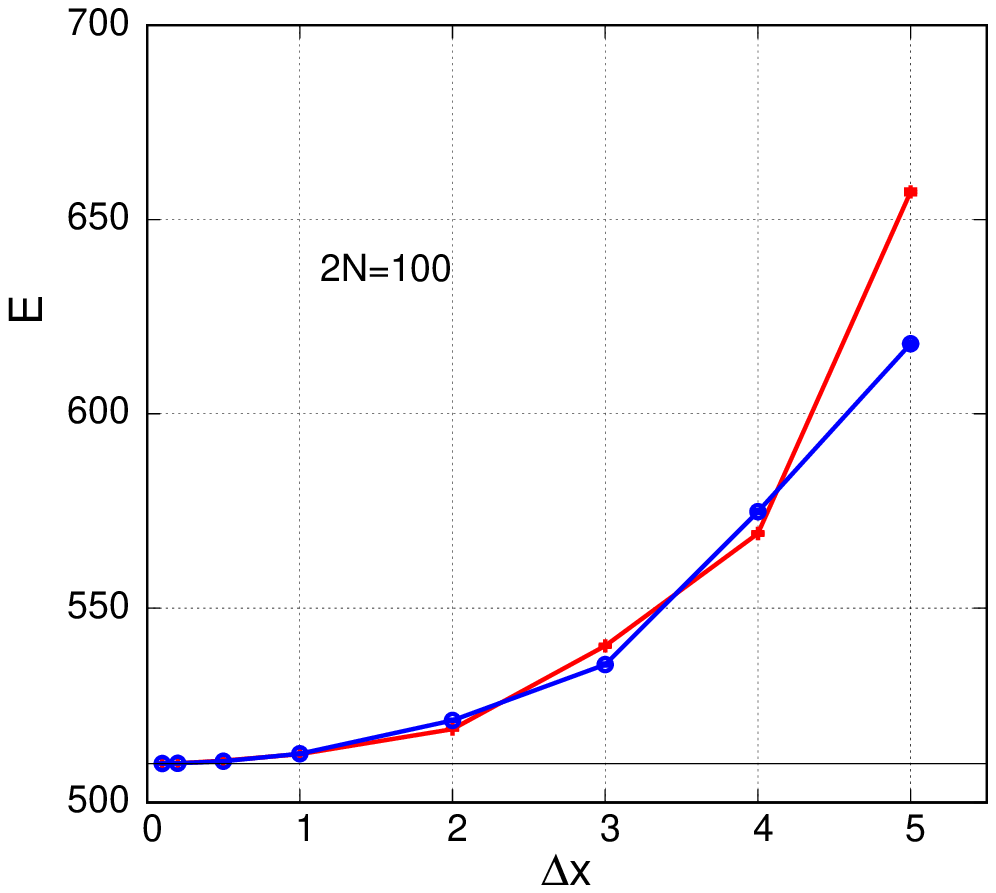}	
	\includegraphics[width=0.48\linewidth]{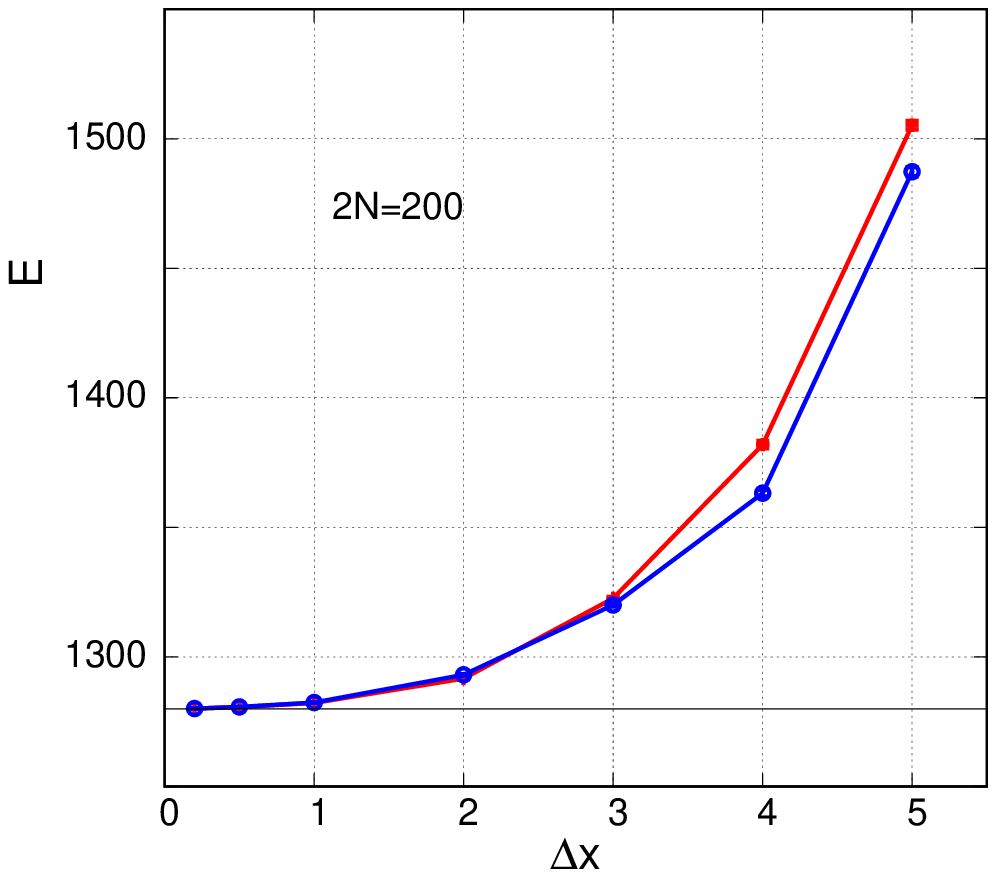}
	\caption{ (color online) The convergence of fermion energy in a 3D harmonic oscillator.
		Exact energies for $2N=$100, 200 spin-balanced fermions are 510 and 1280,
		as shown by the horizontal black line.
		The best calculated results are  510.024$\pm0.003$ and 1280.08$\pm0.01$,
		with systematic errors of 0.005\% and 0.006\%.
	}
	\la{pos200}
\end{figure}

To implement the wave function (\ref{wfd}) in the limit of $\s_i\rightarrow 0$, one generates a set
of random vectors $\s_i=(r_{i1},r_{i2},r_{i3})$ where each component is uniformly distributed 
about the origin as $-\Delta x <r_{ik}<\Delta x$. One then decreases $\Delta x$ to see how the energy
value is approached by $E_{2N}$. This is shown in Fig.\ref{pos200} for $2N=100$ and $200$.
The energy convergence is clearly second-order in $\Delta x$, befitting the fact a non-zero $\Delta x$ 
implies a set of finite $\s_i$, which are
first order errors in the wave function. At large values $\Delta x$, with small number of fermions,
the set of random vector $\s_i$ can clump, resulting in different energy values, as shown in two separate runs
of different random number sequence. When $\Delta x$ is reduced, there is less room for clumping and less scattering
in the energy value. 

\begin{figure}[hbt]
	\includegraphics[width=0.48\linewidth]{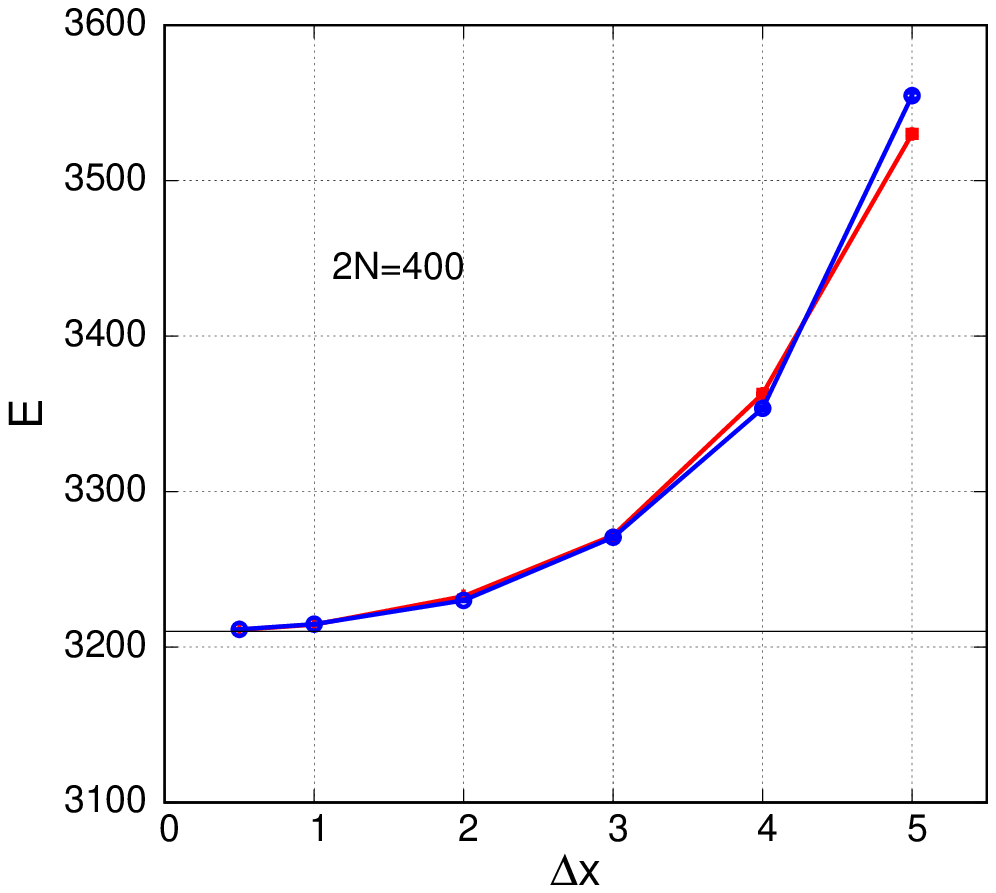}	
	\includegraphics[width=0.48\linewidth]{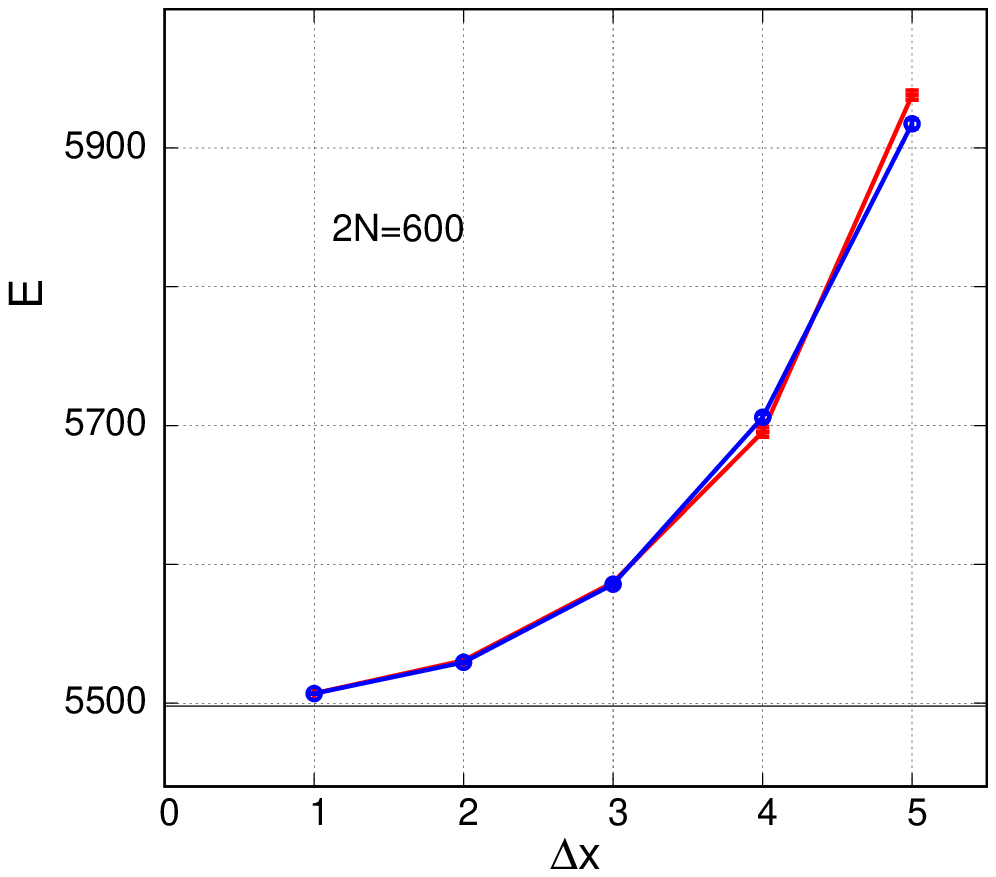}	
	\caption{ (color online)
		Exact energies for $2N$=400 and 600 fermions are 3210 and 5498.
		The best calculated results are 3211.17$\pm 0.04$, 5507.1$\pm 0.2$,
		with systematic errors of 0.03\% and 0.17\%.
	}
	\la{pos400}
\end{figure}

\begin{figure}[hbt]
	\includegraphics[width=0.48\linewidth]{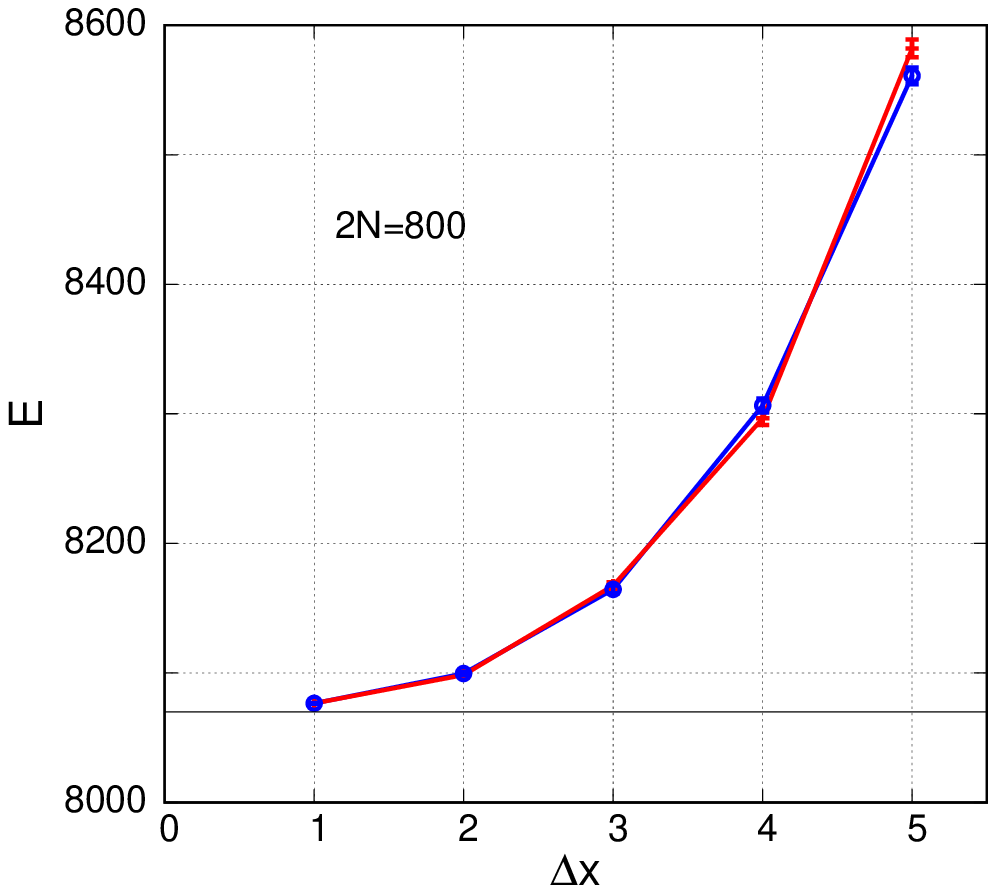}	
	\includegraphics[width=0.48\linewidth]{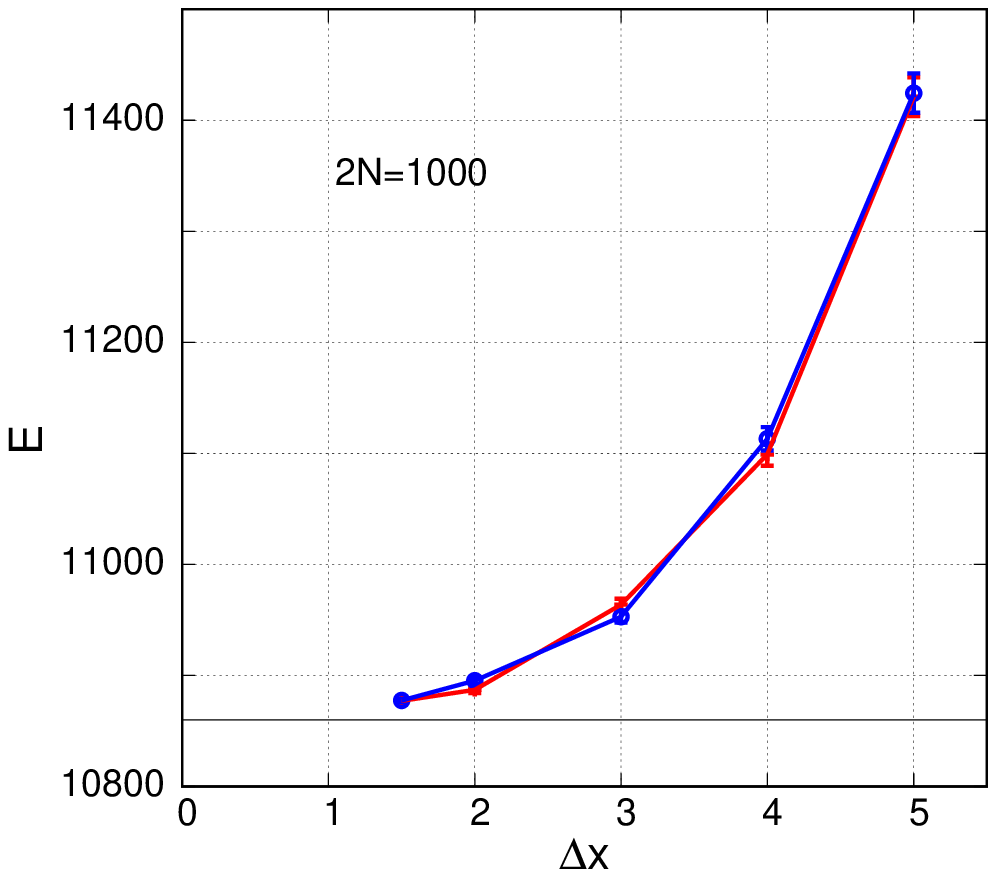}	
	\caption{ (color online)
		Exact energies for $2N$=800 and 1000 are 8070 and 10860.
		The best calculated results are 8076.6$\pm$0.3 and 10877$\pm$1, 
		with systematic errors of 0.08\% and 0.16\%.	
	}
	\la{pos800}
\end{figure}

All particles are moved simultaneous in a single Metropolis trial step; sequential particle moves would
be too slow.  After on the order of $10^5$ Metropolis steps, the statistical errors are very small.
As $\Delta x$ squeezes all $\s_i$ toward zero, the determinant approaches zero, the matrix inversion
subroutine fails and the calculation cannot continue below that value of $\Delta x$. The result is
a systematic error always above the exact energy values, since all $\s_i$ can be viewed as variational
parameters. For $2N=100$ and $200$, this systematic error is too small to be of concern.
The exact energy values are computed in Appendix C.

Fig.\ref{pos400} and Fig.\ref{pos800} shows the results for $2N$=400, 600
and $2N$=800, 1000 respectively. For $2N$=600, 800 and 1000, $\Delta x$ can no longer be reduced 
below one and the systematic error is discernible on the absolute scale. However, on the
relative scale, they remained below 0.2\%. This calculation was done using Fortran with only double-precision.
If quad-precision were used, this systematic error can be further reduced at smaller values of $\Delta x$.

\section {Conclusions}
\la{con}

The thermodynamics of a system of harmonic confined fermions in 3D, including its ground state
energies, has been extensively studies by Brosens {\it et al.}\cite{bro98} using the exact
harmonic propagator in the low temperature limit. In that study, the ground state energy is 
its final goal. Here, we have proposed a remarkably simple wave function (\ref{detwf}) that 
can yield the ground state energy directly, for up to a thousand harmonic fermions,
without evaluating any single particle wave function.
Moreover, because it is a wave function, and not a propagator, it can be part of an initial 
trial function for doing very large scale VMC, or even Ground State Path Integral Monte Carlo\cite{sar00,yan17,chin20} 
calculations on {\it interacting} 3D fermions.

It is known that non-interacting fermions, due to Pauli's exclusion principle, can probe the excited states
of any potential. Lyubartsev\cite{lyu05} has even suggested that by anti-symmetrization, one can also probe
the excited states of an interacting, many-particle system. Mathematically, the wave function (\ref{detwf})
can yield the spectrum of the harmonic oscillator in {\it any} dimension. It seems miraculous that
such a simple wave function can automatically account for the degeneracy of each harmonic state in $D$-dimension,
which is the partition of a non-negative integer $N$ by a sum of $D$ non-negative integers.  

The generalization of (\ref{detwf}) to the Coulomb potential case of $V(r)=-Z/r$ would seem to be
\be
\Psi(\x_1,\x_2\dots\x_N)=\lim_{\s_j\rightarrow 0}\det\Bigl(\exp[-Z|\x_i-\s_j|]\Bigr).
\la{detc}
\ee
The wave function (\ref{detwf}) works for the harmonic oscillator because all of its excited state have the same
Gaussian factor $\e^{-r^2/2}$. For the Coulomb potential, excited states have different Slater 
orbitals $\e^{-Zr/n}$ depending on the principle quantum number $n$. Therefore, (\ref{detc})
cannot possibly be the exact wave function for $N$ non-interacting fermions in a Coulomb potential.
Thus the simple wave function (\ref{detwf}) is unique to the harmonic oscillator.

\appendix

\section {The N-fermion wave function in one dimension}
\la{onedim}

The $N=2$ determinant in 1D can be computed alternatively as
\ba
\det\left(\begin{array}{cc}
	\e^{x_1 s_1} & \e^{x_1 s_2}\\
	\e^{x_2 s_1} & \e^{x_2 s_2}
\end{array}\right)&=&\e^{x_1 s_1}\e^{x_2 s_2}-\e^{x_2 s_1}\e^{x_1 s_2}\nn\\
&=&\sum_{n_1=0}^\infty \frac1{n_1!}x_1^{n_1} s_1^{n_1}\sum_{n_2=0}^\infty \frac1{n_2!}x_2^{n_2} s_2^{n_2}-
\sum_{n_1=0}^\infty \frac1{n_1!}x_2^{n_1} s_1^{n_1}\sum_{n_2=0}^\infty \frac1{n_2!}x_1^{n_2} s_2^{n_2}\nn\\
%&=&\sum_{n_1=0}^\infty\sum_{n_2=0}^\infty\frac1{n_1!n_2!}
%(x_1^{n_1} s_1^{n_1}x_2^{n_2} s_2^{n_2}-x_2^{n_1}s_1^{n_1}x_1^{n_2} s_2^{n_2})\nn\\
&=&\sum_{n_1=0}^\infty\sum_{n_2=0}^\infty\frac1{n_1!n_2!}s_1^{n_1}s_2^{n_2}
(x_1^{n_1}x_2^{n_2}-x_2^{n_1}x_1^{n_2})\nn\\
&=&\sum_{n_1=0}^\infty\sum_{n_2=0}^\infty\frac1{n_1!n_2!}s_1^{n_1}s_2^{n_2}
\det\left(\begin{array}{cc}
	x_1^{n_1} & x_1^{n_2}\\
	x_2^{n_1} & x_2^{n_2} 
\end{array}\right).
\ea
Since $n_1$ and $n_2$ are column indices and the determinant vanishes for
$n_1=n_2$, the sum is only  over
$n_1<n_2$ and $n_2<n_1$. The latter case can be view as the first case with $n_1$ interchanged with $n_2$.
This interchange corresponds to the old determinant with a negative sign, hence
\ba
\det\left(\begin{array}{cc}
	\e^{x_1 s_1} & \e^{x_1 s_2}\\
	\e^{x_2 s_1} & \e^{x_2 s_2}
\end{array}\right)
&=&\sum_{n_1<n_2}\frac1{n_1!n_2!}\left [s_1^{n_1}s_2^{n_2}
\det\left(\begin{array}{cc}
	x_1^{n_1} & x_1^{n_2}\\
	x_2^{n_1} & x_2^{n_2}
\end{array}\right)
-s_1^{n_2}s_2^{n_1}
\det\left(\begin{array}{cc}
	x_1^{n_1} & x_1^{n_2}\\
	x_2^{n_1} & x_2^{n_2}
\end{array}\right)\right ]
\nn\\
&=&\sum_{n_1<n_2}\frac1{n_1!n_2!}
\det\left(\begin{array}{cc}
	s_1^{n_1} & s_1^{n_2}\\
	s_2^{n_1} & s_2^{n_2}
\end{array}\right)
\det\left(\begin{array}{cc}
	x_1^{n_1} & x_1^{n_2}\\
	x_2^{n_1} & x_2^{n_2}
\end{array}\right).
\ea
This expansion was used in the study of the quantum Hall effect by Mikhailov\cite{mik01}.
 
The lowest order terms in $s_i$ are therefore given by $n_1=0$ and $n_2=1$, resulting in
\ba
\det\left(\begin{array}{cc}
	\e^{x_1 s_1} & \e^{x_1 s_2}\\
	\e^{x_2 s_1} & \e^{x_2 s_2}
\end{array}\right)
=
\det\left(\begin{array}{cc}
	1 & s_1\\
	1 & s_2
\end{array}\right)
\det\left(\begin{array}{cc}
	1 & x_1\\
	1 & x_2
\end{array}\right)=s_{21}x_{21},
\la{twobt}
\ea
reproducing (\ref{tpwf}).

Similar manipulations gives the $N=3$ determinant,
\ba
\det\left(\begin{array}{ccc}
	\e^{x_1 s_1} & \e^{x_1 s_2}& \e^{x_1 s_3}\\
	\e^{x_2 s_1} & \e^{x_2 s_2}& \e^{x_2 s_3}\\
	\e^{x_3 s_1} & \e^{x_3 s_2}& \e^{x_3 s_3}
\end{array}\right)
=
\sum_{n_1<n_2<n_3}^\infty\frac1{n_1!n_2!n_3!}
\det\left(\begin{array}{ccc}
	s_1^{n_1} & s_1^{n_2}& s_1^{n_3}\\
	s_2^{n_1} & s_2^{n_2}& s_2^{n_3}\\
	s_3^{n_1} & s_3^{n_2}& s_3^{n_3}
\end{array}\right)
\det\left(\begin{array}{ccc}
	x_1^{n_1} & x_1^{n_2}& x_1^{n_3}\\
	x_2^{n_1} & x_2^{n_2}& x_2^{n_3}\\
	x_3^{n_1} & x_3^{n_2}& x_3^{n_3}
\end{array}\right).\nn\\
\la{three}
\ea
The first non-vanishing term in the above sum, is the third order term in $s_i$ given by $n_1=0$, $n_2=1$, $n_3=2$, 
\ba
&&=\frac1{2}
\det\left(\begin{array}{ccc}
	1 & s_1& s_1^{2}\\
	1 & s_2& s_2^{2}\\
	1 & s_3& s_3^{2}
\end{array}\right)
\det\left(\begin{array}{ccc}
	1 & x_1& x_1^{2}\\
	1 & x_2& x_2^{2}\\
	1 & x_3& x_3^{2}
\end{array}\right)
=\frac12 s_{21}s_{31}s_{32} x_{21}x_{31}x_{32},
\la{threebt}
\ea
which correctly changes sign whenever $x_i\leftrightarrow x_j$.

For $N$-fermion, the above two cases (\ref{twobt}) and (\ref{threebt}) generalize to the
the $N\times N$ Vandermonde determinant:
\be
\det\left(\begin{array}{ccccc}
	1 & x_1 &x_1^2& \cdots &x_1^{N-1}\\
	1 & x_2 &x_2^2& \cdots &x_2^{N-1}\\
	1 & x_2 &x_2^2& \cdots &x_2^{N-1}\\		
	1 & \cdots &\cdots& \cdots &\cdots\\	
	1 & x_N &x_n^2& \cdots &x_N^{N-1}		
\end{array}\right)=\prod_{1\le i<j\le N}(x_j-x_i),
\la{vand}
\ee
yielding the $N$-fermion wave function:
\be
\Psi(x_1,x_2,\cdots x_n)\propto \prod_{i<j} (s_j-s_i)\prod_{i<j} (x_j-x_i)\exp(-\sum_{i=1}^Nx_i^2/2).
\ee
This is an example in which the determinant wave function can be explicit given without evaluating 
any specific single particle wave function, {\it i.e.}, Hermit polynomials.  

\section {The Hamiltonian energy estimator}

Since the below applies equally to $\m_\uparrow$ and $\m_\downarrow$,
for clarity, we will suppress the $\uparrow$ and $\downarrow$ labels.  Given that
\be
{\bf M}=M_{lk}=\exp\left[-\frac1{2}(\x_l-\s_k)^2\right],
\la{mat2}
\ee
one has
\ba
\nabla_i S&=&\nabla_i\ln({\det}{\bf M})
={\rm Tr}[{\bf M}^{-1}\nabla_i{\bf M}]
=\sum_{kl}M^{-1}_{kl}\nabla_i M_{lk},\nn\\
&=&
-\sum_{kl}M^{-1}_{kl}(\x_l-\s_k)\delta_{il}M_{lk}
=-\sum_{k}M^{-1}_{ki}(\x_i-\s_k)M_{ik},\nn\\
&=&-(\x_i-\sum_{k} \s_k M_{ik} M^{-1}_{ki})=-(\x_i-\tilde\s_i),
\ea
and therefore
\ba
\nabla^2_i S&=&-\nabla_i\cdot (\x_i-\sum_{k}\s_k M_{ik}M^{-1}_{ki}),\nn\\
&=&-D+\sum_{k}\s_k\cdot \nabla_i(M_{ik}M^{-1}_{ki}).
\la{b3}
\ea
In the following, repeated indices $l$ and $n$ are summed over, but not for $i$ or $k$,
\ba
\nabla_i (M_{ik}M^{-1}_{ki})&=&
-(\x_i-\s_k)M_{ik}M^{-1}_{ki}
-M_{ik}M^{-1}_{kl}(\nabla_iM_{ln})M^{-1}_{ni}\nn\\
&=&
-\Bigl[(\x_i-\s_k)M_{ik}M^{-1}_{ki}
-M_{ik}M^{-1}_{ki}(\x_i-\s_n)M_{in}M^{-1}_{ni}\Bigr]\nn\\
&=&-\Bigl[\x_iM_{ik}M^{-1}_{ki}-\s_k M_{ik}M^{-1}_{ki}-M_{ik}M^{-1}_{ki}\x_i
+M_{ik}M^{-1}_{ki}\s_nM_{in}M^{-1}_{ni}  \Bigr]\nn\\
&=&-\Bigl[-\s_k M_{ik}M^{-1}_{ki} +M_{ik}M^{-1}_{ki}\tilde\s_i  \Bigr].
\ea
Now the sum over $k$ in (\ref{b3}) yields
\ba
\sum_{k}\s_k\cdot \nabla_i(M_{ik}M^{-1}_{ki})&=&-\sum_{k}\s_k\cdot
\Bigl[-\s_k M_{ik}M^{-1}_{ki} +M_{ik}M^{-1}_{ki}\tilde\s_i  \Bigr]
\nn\\
&=&\sum_{k}\s^2_k M_{ik}M^{-1}_{ki}-\tilde\s_i^2,
\ea
and the final sum over $i$ gives
\ba
\sum_{i=1}^N\nabla^2_i S=-ND+\sum_{i=1}^N(\s_i^2-\tilde\s_i^2 ).
\ea
The local energy is therefore
\ba
E_L
&=&\sum_{i=1}^N\left[-\frac12[\nabla^2_i S_{\uparrow}+(\nabla_i S_{\uparrow})^2]+\frac12 \x_i^2\right],\nn\\
&=&N\frac{D}2-\frac12\sum_{i=1}^N(\s_i^2-\tilde\s_i^2 )
-\frac12\sum_{i=1}^N(\x_i-\tilde\s_i)^2 + \frac12\sum_{i=1}^N\x_i^2.
\ea
If $\m$ were diagonal, as in the boson case, then $\tilde\s_i=\s_i$, and the above is just
\ba
E_L
&=&N\frac{D}2
-\frac12\sum_{i=1}^N(\x_i-\s_i)^2 + \frac12\sum_{i=1}^N\x_i^2,
\ea
which correctly reproduces the $N$-boson energy of $ND/2$ when $\s_i\rightarrow 0$.

\section{Fermion energies in a 3D harmonic oscillator} 

The spectrum of the 3D harmonic oscillator is given by
\be
E_m=(m-1)+\frac32=\frac{2m+1}2,
\ee
where 
\be 
m-1=n_x+n_y+n_z,
\ee 
with the ground state corresponding to $m=1$.

Given $m$, $n_x$ can take on $m$ values from 0 to $(m-1)$. 
For $n_x=(m-1)$, one must have just $(n_y,n_z)=(0,0)$. For $n_x=(m-1)-1$,
one can have $(n_y,n_z)=$(1,0) or (0,1). For $n_x=(m-1)-2$, $(n_y,n_z)=(2,0), (1,1), (0,2)$, etc..
The total degeneracy for level $E_m$ is therefore $1+2+3+\cdots+m$, given by 
\be
g=\frac12 m(m+1).
\ee
Therefore, for closed-shell occupation up to and including the $n^{th}$ level, 
we have total particle number and energy
$$
N(n)=\sum_{m=1}^{n}\frac12 m(m+1),
$$
$$
E(n)=\sum_{m=1}^{n}\frac12 m(m+1)\frac{2m+1}2.
$$
For equal spin-up and spin-down fermions, multiply by 2 gives
$$
N(n)=\sum_{m=1}^{n}m(m+1),
$$
$$
E(n)=\sum_{m=1}^{n}\frac12 m(m+1)(2m+1).
$$
Since we have the power sums
\ba
\sum_{m=1}^{n}m&=&\frac12 n(n+1),\\
\sum_{m=1}^{n}m^2&=&\frac16 n(n+1)(2n+1),\\
\sum_{m=1}^{n}m^3&=&\frac14n^2(n+1)^2,
\ea
we have the following {\it closed shell} results
\ba
N(n)&=&\frac12 n(n+1)+\frac16 n(n+1)(2n+1)=\frac13 n(n+1)(n+2),\\
E(n)&=&\frac14 n(n+1)[n^2+3n+2]=\frac14 n(n+1)^2(n+2),
\ea
which agree with a previous, different derivation of Brosens {\it et al.}\cite{bro98} where
their $L=n-1$. 

For any $M$ between two closed shells $n$ and $n+1$, $N(n)\le M\le N(n+1)$, the energy is
$E(n)$ plus the number of particle greater then $N(n)$, which is $[M-N(n)]$, times
the energy level at $m=n+1$, which is
\be
E_M=E(n)+[M-N(n)]\left( \frac{2(n+1)+1}2 \right).
\ee

%%%%%%%%%%%%%%%%%%%%%%%%%%%%%%%%%%%%%%%%%%%%%%%%%%%%%%%

\newpage
\begin{thebibliography}{99}
	
\bibitem{cep96} D. M. Ceperley
{\it Path Integral Monte Carlo Methods for Fermions} ed K Binder and G Cicotti (Bologna:
Editrice Compositori) Simulation in condensed matter, physics and chemistry, 1996.
	

\bibitem{ass07} R. Assaraf, M.  Caffarel and A. Khelif,
%	`` The fermion Monte Carlo revisited ",
J. Phys. A: Math. Theor. 40(2007) 1181–1214	

	
\bibitem{sar00} A. Sara, K. E. Schmidt and and W. R. Magro,
%	``A path integral ground state method." 
	J. Chem. Phys. {\bf 113}, 1366-1371 (2000).
	
\bibitem{yan17} Y. Yan and D. Blume,
%	``Path integral Monte Carlo ground state approach: Formalism, implementation, and applications",
	J. Phys. B: At. Mol. Opt. Phys. 50, 223001 (2017).
	
\bibitem{chin20}S. A. Chin,
%Solving fermion problems without solving the sign problem: Symmetry-breaking wave functions
%from similarity-transformed propagators for solving two-dimensional quantum dots
Phys. Rev. E 101, 043304 (2020).	

\bibitem{yos79} D. Yoshioka, H. Fukuyama, J. Phys. Soc. Japan 47 (1979) 394.
\bibitem{yos83} D. Yoshioka, P.A. Lee, Phys. Rev. B 27 (1983) 4986.
\bibitem{mak83} K. Maki, X. Zotos, Phys. Rev. B 28 (1983) 4349.
\bibitem{lau83} R. B. Laughlin, Phys. Rev. Lett. 50 (1983) 1395.
\bibitem{kai02} J. Kainz, S. A. Mikhailov, A. Wensauer, and U. R\"ossler,
                 Phys. Rev. B 65, 115305 (2002).
\bibitem{har02} A. Harju, S. Siljamaki, and R. M. Nieminen,
%Wigner molecules in quantum dots: A quantum Monte Carlo study
                Phys. Rev. B 65, 075309 (2002)

\bibitem{yan99} C. Yannouleas and U. Landman
 %Spontaneous Symmetry Breaking in Single and Molecular Quantum Dots
 Phys. Rev. Lett. 82, 5325 (1999)– Erratum Phys. Rev. Lett. 85, 2220 (2000)



\bibitem{chin90} Siu A. Chin, 
%``Quadratic diffusion Monte Carlo algorithms for solving atomic many-body problems'', 
Phys. Rev. A 42, 6991 (1990).

\bibitem{fey72}
R. P. Feynman, {\it Statistical Mechanics - A Set of Lectures} 
(Benjamin Advanced Book, Reading, MA, 1972).

\bibitem{uhl} G. E. Uhlenbeck and L. S. Ornstein, 
 %"On the theory of Brownian Motion". 
 Phys. Rev. 36,  823–841 (1930)


\bibitem{caf88} M. Caffarel and P. Claverie,
J. Chem. Phys. 88, 1088 (1988); 88, 1100 (1988).

\bibitem{mos82} J. W. Moskowitz, K. E. Schmidt, M. E. Lee, and M. H. Kalos,
J. Chem. Phys. 77, 349 (1982).

\bibitem{rey82}P. J. Reynolds, D. M. Ceperley, B. J. Adler, and W. A. Lester,
J. Chem. Phys. 77, 5593 (1982).

\bibitem{met} N. Metropolis, A. W. Rosenbluth, M. N. Rosenbluth, A. H. Teller and E. Teller, 
%"Equations of State Calculations by Fast Computing Machines". 
J. Chem. Phys. 21, 1087 (1953)

\bibitem{mik01} S. A. Mikhailov,
%A new approach to the ground state of quantum Hall systems.
%Basic principles
Physica B 299, 6 (2001).

%Confined harmonically interacting spin-polarized fermions
\bibitem{bro98} F. Brosens, J. T. Devreese and L. F. Lemmens, 
Phys. Rev. E {\bf 57}, 3871 (1998).

\bibitem{lyu05} A. P. Lyubartsev,
J. Phys. A: Math. Gen. 38 (2005) 6659–6674.	







% -----------------------------end----------------------------------



\end{thebibliography}
\end{document}